\documentclass{ptephy_v1}

\usepackage{graphicx}
\usepackage{epsfig}
\usepackage{tabularx,ragged2e,booktabs,caption}
\usepackage{dcolumn}
\usepackage{bm}
\usepackage{amssymb}
\usepackage{calrsfs}
\DeclareMathAlphabet{\pazocal}{OMS}{zplm}{m}{n}
\usepackage{amsmath}
\usepackage{subfigure}

\newcommand{\be}{\begin{equation}}
\newcommand{\ee}{\end{equation}}
\newcommand{\bea}{\begin{eqnarray}}
\newcommand{\eea}{\end{eqnarray}}

\newcommand{\tPsi}{{\tilde \Psi}}
\newcommand{\tpsi}{{\tilde \psi}}

\newcommand{\pro}{\partial}

\newcommand{\ba}{\begin{array}}
\newcommand{\ea}{\end{array}}

\newcommand{\nn}{\nonumber}

\newcommand{\Da}{\mathcal{D}}

\newcommand{\Ba}{\mathcal{B}}

\begin{document}

\title{Stable spherically symmetric monopole field background in a pure QCD}


\author{Youngman Kim}
\affil{Rare Isotope Science Project, Institute for Basic Science, 
Daejeon 305-811, Korea \email{ykim@ibs.re.kr}}

\author{Bum-Hoon Lee}
\author[3]{}
\affil{Asia Pacific Center of Theoretical Physics, 
Pohang 790-330, Korea}

\author{D.G. Pak}
\author[2]{} 
\affil{CQUEST, Sogang University, Seoul 121-742, Korea \email{bhl@sogang.ac.kr}}
\author[4]{}
\affil{Chern Institute of Mathematics, Nankai University, 
Tianjin 300071, China \email{dmipak@gmail.com}}

\author{Takuya Tsukioka}
\affil{School of Education,\\ Bukkyo University, Kyoto 603-8301,
Japan \email{tsukioka@bukkyo-u.ac.jp}}


\begin{abstract}%
We consider a stationary spherically symmetric monopole like solution 
with a finite energy density in a pure quantum chromodynamics (QCD). 
The solution can be treated as a static Wu-Yang monopole dressed 
in time dependent field corresponding to off-diagonal gluons. 
We have proved that such a stationary monopole field represents 
a background vacuum field of the QCD effective action 
which is stable against quantum gluon fluctuations. 
This resolves a long-standing problem of existence of a stable vacuum 
field in QCD and opens a new avenue towards microscopic theory of the vacuum.
\end{abstract}

\subjectindex{A34, B03, B35}

\maketitle

\section{Introduction}

Non-perturbative structure of the vacuum and
confinement phenomenon represent two most important and closely related problems
in foundations of QCD which is supposed to be a basic fundamental theory
of strong interactions. Despite on significant progress 
in lattice studies \cite{Phil2008} 
there is still no deep knowledge about the microscopic vacuum structure
and origin of the color confinement from the first principles of QCD 
(see \cite{diGiacomo2014} and references therein).
One of the attractive mechanisms of quark confinement is based on the
Meissner effect in dual color superconductor which assumes
generation of the monopole condensation
\cite{nambu74,mandelstam76,polyakov77,thooft81,Nair1985}.
The old known Savvidy-Nielsen-Olesen QCD vacuum
based on homogeneous magnetic field configuration
\cite{savv,N-O} suffers instability against the quantum fluctuations.
The most popular Copenhagen ``spaghetti'' vacuum 
model \cite{niel-nino,niel-oles,amb-oles1} based on vortex domain 
structure does not provide a consistent microscopic description of
the vacuum.
Numerous attempts to construct a rigorous true vacuum of QCD
using various vacuum field configurations 
(instantons \cite{shuryak1997}, monopoles,
dyons \cite{1995simonov,kumar2010}, center vortices \cite{engelhardt2000},
monopoles \cite{choprl80,chopakprd2002,diGiacomo2015,pak05}, 
etc. \cite{schan82,bordag}) show that the existence of a stable vacuum 
represents a most serious long-standing problem 
in QCD. One of principal obstacles in resolving that problem was absence of a
proper regular classical solution which
must be stable against vacuum fluctuations at microscopic space scale and which
can serve as a structure element in formation of the vacuum.
Note that all previous vacuum models employ static vacuum field configurations
none of which do not possess quantum stability at microscopic scale. For instance, 
a single static vortex field in the ``spaghetti'' vacuum model is
unstable, and the vacuum stability is restored only due to averaging 
procedure over the statistical ensemble of the vortex domains. 
An advantage of the ``spaghetti'' vacuum is that it provides approximate
qualitative description of the vacuum structure, in particular, 
it was conjectured from quantum 
mechanical consideration that color magnetic vortices should vibrate 
at small space-time scale. 
This gives a hint that time dependent field configurations might play 
an important role in vacuum structure.

In the present paper we explore an idea that stationary generalized 
monopole like solution can serve as an initial structure element 
in formation of the QCD vacuum.
We consider a recently proposed classical stationary spherically
symmetric monopole solution which can be treated as a system 
of a static Wu-Yang monopole interacting to time-dependent 
off-diagonal gluon field \cite{pakp1}.
We have proved that such a solution provides a monopole vacuum field 
which is stable against gluon fluctuations in one-loop approximation. 
This opens a new way towards construction of a microscopic theory 
of the QCD vacuum through the condensation of monopole and/or 
monopole-antimonopole pairs.

\section{Stationary generalized Wu-Yang monopole solution}

Let us describe the main properties of the spherically symmetric 
stationary monopole solution proposed in \cite{pakp1}. 
Such a solution will be used in the subsequent sections 
as a background field in the effective action functional, 
and its non-trivial properties will provide the stability under 
quantum gluon fluctuations.
We start with a standard classical Lagrangian of a pure  $SU(3)$ QCD
\be
{\cal L}=-\dfrac{1}{4}F_{a\mu\nu}F^{a\mu\nu}. \label{Lagr0}
\ee
We introduce the following spherically symmetric ansatz
for non-vanishing components of the gauge potential
in spherical coordinates  $(r, \theta, \varphi)$:
\be
\begin{array}{rcl}
A_\varphi^1\!\!&=&\!\! -\dfrac{1}{r}\psi_1(t, r), \ 
A_\theta^2=\dfrac{1}{r}\psi_1 (t, r), \ 
A_\varphi^3= \dfrac{1}{gr} \cot \theta, 
\\
A_\varphi^4\!\!&=&\!\!\dfrac{1}{r}\psi_2(t, r), \
A_\theta^5=\dfrac{1}{r}\psi_2 (t, r), \ 
A_\varphi^8=-\dfrac{\sqrt 3}{gr} \cot \theta, 
\end{array}
\label{spherwav} 
\ee
where Abelian gauge potentials 
$A_{\varphi}^3$ and $A_{\varphi}^8$ describe a static Wu-Yang monopole
with a total color magnetic charge two, $g_m^{\rm tot}=2$, \cite{choprl80},
and the functions $\psi_{1}(t,r)$ and $\psi_{2}(t,r)$ correspond to 
dynamical degrees of freedom 
of the off-diagonal components of the gluon field $A_\mu^a$.
The ansatz (\ref{spherwav}) describes two coupled monopole field
configurations corresponding to $I$- and $U$-type  subgroups $SU(2)$. 
One can verify that the ansatz  (\ref{spherwav})  is consistent with all equations of motion
of the pure $SU(3)$ QCD,
and substitution of the ansatz into the equations of motion
results in only two non-trivial  independent partial differential equations
\be
\begin{array}{rcl}
&&\pro^2_{t} \psi_1 -\pro^2_r \psi_1+\dfrac{g^2}{2 r^2} \psi_1
\Big(2 \psi_1^2-\psi_2^2-\dfrac{2}{g^2}\Big)=0, \\
&&\pro^2_{t} \psi_2 -\pro^2_r \psi_2+\dfrac{g^2}{2 r^2} \psi_2
\Big(2
 \psi_2^2-\psi_1^2-\dfrac{2}{g^2}\Big)=0.
\end{array}
\label{eqsu3}
\ee
In a special case when  $A_\theta^5=A_\varphi^4=A_\varphi^8= 0$
the ansatz describes a $SU(2)$ embedded field configuration which
corresponds to a system of the Wu-Yang monopole with a
unit monopole charge, $g_m^{\rm tot}=1$, interacting to off-diagonal gluon field.
With $\psi_1(t, r)\equiv \psi(t, r),~ \psi_2(t,r)=0$ the equations (\ref{eqsu3})
reduce to one differential equation
 \be
\pro^2_{t} \psi -\pro^2_r \psi+\dfrac{g^2}{r^2} \psi
\Big(\psi^2-\dfrac{1}{g^2}\Big)=0.  
\label{eqsu2}
\ee

The equation (\ref{eqsu2}) admits a wide class of time dependent
solutions including non-stationary soliton like propagating solutions 
in the effective two-dimensional space-time 
$(r,t)$ \cite{p14,p15,p16,p17,p18}. 
We show that there is a subclass of generalized stationary Wu-Yang monopole
solutions which possess a finite energy density. The most important
issue is that such solutions are stable against quantum gluon
fluctuations 
in pure QCD. 
For simplicity we consider first the vacuum stability problem 
in the case of $SU(2)$ Yang-Mills theory.
 In that case by performing an appropriate gauge transformation
\cite{choprd80} one can rewrite 
the $SU(2)$ part of the ansatz (\ref{spherwav}) in a regular gauge as follows
($a=1,2,3$)
\be
A_m^a= -\epsilon^{abc}\hat n^b \pro_m \hat n^c \Big (\dfrac{1}{g}-\psi(t,r) \Big), \label{wuyang}
\ee
where $\hat n=\vec r/r$. 
It is clear that the ansatz (\ref{wuyang})
describes a generalized time dependent Wu-Yang monopole field configuration.
A known static Wu-Yang monopole corresponds to the limiting case $\psi(t,r)=0$,
and a trivial pure gauge vacuum configuration is described 
by $\psi(t,r)= \pm 1/g$.
We prefer the ansatz written in the so-called singular 
Abelian gauge \cite{choprd80}, (\ref{spherwav}), 
since such a notation is more suitable for description of stationary
monopole solutions in $SU(N)$ Yang-Mills theory and in description of 
multimonopole configurations. 

For simplicity we consider first the vacuum stability problem 
in the case of $SU(2)$ embedded solution.
For arbitrary function $\psi(t,r)$ one has 
the following non-vanishing field strength components
\be
\begin{array}{rcl}
F_{t\varphi}^1\!\!&=&\!\!-\dfrac{1}{r}\pro_t\psi, \ \ 
F_{r\varphi}^1=-\dfrac{1}{r}\pro_r \psi, \ \ 
\\
F_{t\theta}^2\!\!&=&\!\!\dfrac{1}{r}\pro_t \psi, \ \ 
F_{r\theta}^2=\dfrac{1}{r}\pro_r \psi, \ \ 
F_{\theta\varphi}^3=\dfrac{1}{gr^2}(g^2\psi^2-1). 
\end{array}
\ee
The radial magnetic field component $F_{\theta\varphi}^3$
generates a non-zero magnetic flux through a sphere with a center
at the origin, $r=0$.
So that, the color magnetic charge of the monopole depends 
on time and distance from the center. 
Note that various generalized static Wu-Yang monopoles have been 
considered before, all of them have singularities in agreement with the
Derrick's theorem \cite{derr}.  
Note that presence of singularities in the expressions for 
the gauge potential (\ref{spherwav})  represents coordinate
singularities related to the 
chosen singular gauge. Such coordinate singularities disappear in the 
regular gauge (\ref{wuyang}). 
By substitution of the ansatz  (\ref{spherwav}) into the energy functional 
one can verify that the energy density is finite everywhere
\be
E=
4\pi\!\!\int\! {\rm d}r\, 
\Big ((\pro_t \psi)^2+(\pro_r \psi)^2 
+\dfrac{1}{2g^2r^2}(g^2 \psi^2-1)^2 \Big ) .
\label{entotspher}
\ee

The equation (\ref{eqsu2}) admits a local non-static solution near the origin
which removes the singularity of the monopole at the center 
\bea
\psi(t,r)&=&\dfrac{1}{g}+\sum_{n=1} c_{2 n}(t) r^{2 n},  \nn \\
c_4(t)&=& \dfrac{1}{10} \big (3 g c_2^2(t) +c_2''(t)\big ),   \nn \\
 c_6(t)&=&\frac{1}{28}\big(c_4''(t)+6gc_2(t)c_4(t)+g^2c_2^3(t)\big), \nn \\
&\vdots&       \label{locsol}
\eea
where the coefficient functions $c_{2n}(t)$ ($n\geq 2$) are determined in 
terms of one arbitrary function $c_2(t)$. 
 In asymptotic region, $r\simeq \infty$,
the non-linear 
equation (\ref{eqsu2}) reduces to a free D'Alembert
equation which has a standing spherical wave solution
\bea
\psi(t, r) &\simeq& a_0+A_0 \cos (M r) \sin(M t)
+{\mathcal O}\Big(\dfrac{1}{r}\Big), \label{asymsol}
\eea
where $a_0, A_0$ are integration constants,
the mass scale $M$ appears due to scaling invariance 
in the theory under dilatations $r \rightarrow Mr, t\rightarrow Mt $. 

\begin{figure}[!h]
\centering
\includegraphics[width=70mm,height=52mm, bb=0 0 562 377]{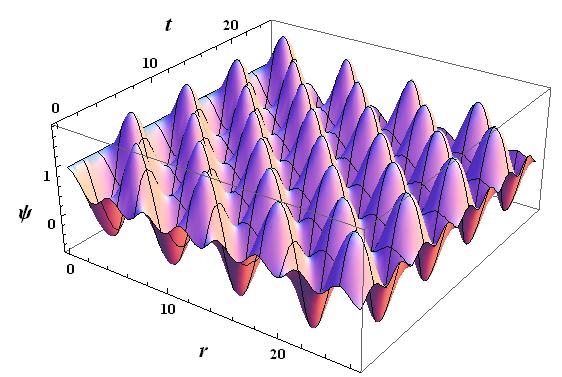}
\caption[fig1]{Stationary monopole solution;
$(L_0\leq r,t \leq L)$, $L_0=10^{-6},~L=8 \pi$, $c_0=-0.041$,~$c_1=-0.523,~g=1$
.}\label{Fig1}
\end{figure}

To solve numerically the Eq.  (\ref{eqsu2}) 
we use the local solution  (\ref{locsol}) 
to impose initial Dirichlet conditions 
along the boundary $r=L_0$ in the numeric domain
$(L_0\leq r \leq L,~ 0\leq t \leq L)$. 
The initial profile function $c_2(t)$ can be chosen arbitrarily as any regular
periodic function, we set
\begin{equation}
 c_2(t)=c_0+c_1 \sin (M t),  \label{locsol10}
\end{equation}
where $c_0, \, c_1$ are numeric constants. 
Note that only one of two parameters in the local solution (\ref{locsol10})
is free, the other is fixed by the requirement that a
numeric solution matches the asymptotic solution (\ref{asymsol}).
Dimensional analysis implies that the energy of the solutions
is proportional to $M$ so that the energy vanishes in the limit $M\rightarrow 0$.
This might cause some doubts on stability of the solution.
However, one should stress that standard arguments
on existence of static solitonic solutions based on the Derrick's theorem
\cite{derr} are not applicable to the case of stationary solutions 
which satisfy a condition of extremum value of the classical action, 
not the energy functional. In  the case of Yang-Mills theory 
the action is conformal invariant
and its first variational derivative with respect to the scale parameter 
$M$ vanishes identically. 
So the parameter $M$ determines the scale of the space-time coordinates
and can be set to one without loss of generality.
With this one can solve numerically the equation (\ref{eqsu2}), and
the corresponding solution is depicted in Fig.\ref{Fig1}.
Note that one has a stiffness numeric problem near the origin, so that  
we have checked the regular structure and convergence of the numeric solution 
in close vicinity of the origin up to $L_0=10^{-6}$ while keeping the 
radial size of the numeric domain up to $L=64 \pi$. 

The solution has several surprising features.
The field configuration of the solution includes a static
Wu-Yang monopole immersed in a standing spherical wave made of
off-diagonal gluons.
The standing wave does not screen completely the color monopole
charge at large distances.  
One can find that in the asymptotic region the function $\psi(t,r)$
oscillates around the value $a_0\simeq 0.65$.
So the radial component of the color magnetic field
$F_{\theta \varphi}^1$ has a non-zero averaged value
which provides a non-vanishing total color magnetic charge.
Another interesting feature of the solution is that a corresponding 
canonical spin density
vanishes identically. This gives a hint that such a solution,
treated as a quantum mechanical wave function, can lead to 
a stable condensate of massive spinless particles. 
An idea that particles (or pseudo-particles) can be described by 
stationary solutions was sounded long time ago \cite{derr, jackiw77,jackiwRMP}.
The main obstacle towards practical realization of this idea 
was absence of known regular stationary solutions in real QCD.
Even though our solution has an infinite total energy, the energy 
density falls down as $1/r^2$  with increasing the distance 
from the center of the monopole.
Note that our solution differs from the known non-linear spherical 
wave type solutions which
have singularity at the center and light speed velocity. 
The most important question related to properties of the solutions
is whether the monopole solution is stable against the quantum
gluon fluctuations.

\section{Vacuum stability of the monopole field in $SU(2)$ QCD}

\subsection{Effective action approach to vacuum stability problem}

In order to prove the quantum stability of the monopole field 
we study the structure of the effective action of QCD using
the standard functional integral methods. We will consider the stability of the monopole
solution under small quantum fluctuations of the gluon field. 
An initial gauge potential $A_\mu^a$ is split into a classical $\Ba_\mu^a$,
and a quantum $Q_\mu^a$ parts
\be
A_\mu^a=\Ba_\mu^a+Q_\mu^a.\label{splitpot}
\ee
We choose a temporal gauge for the classical 
and the quantum field, $\Ba_t^a=Q^a_t=0$. The temporal gauge
has a residual symmetry which can be fixed by
imposing an additional covariant Coulomb gauge condition, ${\Da}_i
Q^{ia}=0$,
with a covariant derivative ${\Da}_i$ containing the background
field $\Ba_i^a$.
Hereafter we use the Latin indices for the spacial components.
A one-loop correction to the classical action
is obtained by functional integration over
the quantum fields $Q_i^a$, and it can be expressed
in terms of functional logarithms in  a compact form
\cite{yildiz80,claudson80,adler81,dittrich83,flory83,blau91,
reuter97,chopakprd2002}
\bea
S^{\rm 1 loop}_{\rm eff}&=&-\dfrac{1}{2} {\rm Tr} \ln [K_{ij}^{ab}]+
{\rm Tr}\ln [M_{\rm FP}^{ab}], \nn \\
K_{ij}^{ab}&=&-
\delta^{ab} \delta_{ij} \pro^2_t
            -\delta_{ij}({\Da}_k {\Da}^k)^{ab} -2 f^{acb}{\mathcal
	    F}_{ijc}, \quad \label{Koper}\\
M_{\rm FP}^{ab}&=&-({\Da}_k{\Da}^k)^{ab},   \nn 
\eea
where the operators $K_{ij}^{ab}$ and 
$M^{ab}_{\rm FP}$ correspond to the contributions of
gluons and Faddeev-Popov ghosts, respectively.
The operator $K_{ij}^{ab}$ represents a 
well-defined second order differential operator
of elliptic type in four-dimensional Euclidean space $(t, r,\theta,\varphi)$.
The question of vacuum stability is reduced to the question of
whether the operator $K_{ij}^{ab}$ has negative
eigenvalues for a given classical background field $\Ba_i^a$.
To find the eigenvalues one has to solve the Schr\"{o}dinger type
equation with the operator $K_{ij}^{ab}$ treated as a Hamiltonian
operator for a quantum mechanical system
\be
K_{ij}^{ab} \Psi_j^b=\lambda \Psi_i^a, \label{schr3}
\ee
where the ``wave functions'' $\Psi_i^a(t, r,\theta,\varphi)$ describe quantum
gluon fluctuations.
Note that the ghost operator originates from the interaction of 
spinless ghost fields with the color magnetic field. Such interaction
does not produce negative tachyon modes \cite{N-O}, so it is sufficient
to study the eigenvalue spectrum of the operator $K_{ij}^{ab}$ only.

One should note that expression  (\ref{Koper}) for the operator $K_{ij}^{ab}$ 
is valid for an arbitrary space-time dependent background field $\Ba_\mu^a$.
Moreover, the equation (\ref{Koper}) is manifestly gauge invariant. 
This follows from the gauge invariant background field method 
applied to one-loop quantization \cite{abbott,siegel}.
A key idea in the background field quantization scheme is that splitting
the initial gauge potential into a sum of the classical, $\Ba_\mu^a$,
and quantum, $Q_\mu^a$,  fields implies two types of symmetries
originating from the original $SU(2)$ gauge transformation.
The first one, (I), is defined by a quantum gauge transformation
\bea
\delta Q_\mu^a&=& (D_\mu \vec \alpha)^a, \nn \\
\delta \Ba_\mu^a&=&0,   \label{typeIgtr}
\eea
where $D_\mu=\pro_\mu +\vec A_\mu$ is a covariant derivative containing the 
full gauge potential, and $\alpha^a(x)$ is an infinitesimal gauge parameter.
The second type of symmetry, (II), is defined by a background gauge transformation
\bea
\delta Q_\mu^a&=& f^{abc} Q_\mu^b \alpha^c, \nn \\
\delta  \Ba_\mu^a&=& ({\Da}_\mu \vec \alpha )^a.  \label{typeIIgtr}
\eea
As a result within the framework of the background quantization 
the classical field $\Ba_\mu^a$ appears in the expression for the
operator $K_{ij}^{ab}$ only in terms of background covariant derivatives and 
gauge field strength ${\mathcal F}_\mu^{ab}$. It is clear that
operator $K_{ij}^{ab}$ transforms as an adjoint vector 
under $SU(2)$ type (II) transformations,
and the fluon fluctuation field $\Psi_i^a$ transforms as a vector 
in fundamental representation.
Taking in account that $SU(2)$ group is locally isomorphic 
to the orthogonal group $SO(3)$ one concludes that the eigenvalue 
spectrum is gauge invariant and does not depend on a 
chosen gauge of the background field.

\subsection{Qualitative analysis of stability of the monopole field}

Before solving the Schr\"{o}dinger equation (\ref{schr3}) numerically,
let us perform a qualitative estimate of ground state eigenvalues
to trace the origin of positiveness of the operator $K_{ij}^{ab}$.
The operator $K_{ij}^{ab}$ does not have explicit dependence
on azimuthal angle $\varphi$, so that the ground state
eigenfunctions $\Psi_i^a$ depend only on three coordinates $(t, r,\theta)$. 
First, we apply the variational method to reduce the three-dimensional
equation (\ref{schr3}) to an effective equation 
in two-dimensional space-time $(t,r)$.
Within the variational approach one has to minimize
the following ``Hamiltonian'' functional
\be
\langle \mathcal{H}\rangle
=\int {\rm d}r\, {\rm d}\theta\,
{\rm d}\varphi\, r^2 \sin \theta\, \Psi_i^a K_{ij}^{ab} \Psi_j^b. \label{funct}
\ee
One can make qualitative estimates assuming
that all ground state eigenfunctions $\Psi_i^a$
includes angle dependence which guarantee the finiteness
of the Hamiltonian
\be
\Psi_i^a(t, r,\theta,\varphi)=f_i^a(t, r) \sin \theta. \label{sufcond2}
\ee
With this one can perform integration over the angle variables $(\theta, \varphi)$
in (\ref{funct}) and obtain an effective Schr\"{o}dinger equation for the ground state
\be
\tilde K_{ij}^{ab} f_j^b(t,r)=\lambda f_i^a(t,r), \label{schreff}
\ee
where the operator $\tilde K_{ij}^{ab}$
includes dependence only on two coordinates $(t, r)$,
\be
\tilde K_{ij}^{ab}=
\delta_{ij}\delta^{ab}\Big(-\pro^2_t-\pro^2_r-\dfrac{2}{r}\pro_r\Big)
+V_{ij}^{ab}(t,r).
\ee
The quadratic form corresponding to the potential $V_{ij}^{ab}$
can be written in the form
\be
f_i^a V_{ij}^{ab} f_j^b = \dfrac{1}{r^2} V_0[f]+\dfrac{1}{r^2} V_1[f]
+\dfrac{\psi^2}{r^2}V_2[f]+\dfrac{\psi}{r^2}V_3[f]+\dfrac{\pro_r
\psi}{r}V_4[f],
\quad \label{quadV}
\ee
where the first term includes contribution from a free vector Laplace operator,
$V_1[f]$ corresponds to contribution of the Wu-Yang monopole 
and $V_{2,3,4}[f]$ contains 
contributions proportional to $\psi^2, \psi$ and $\pro_r \psi$, respectively.
One can verify by using variational minimization procedure 
that a total coefficient function in front of the centrifugal 
potential $1/r^2$ is positively defined for arbitrary
fluctuating functions $f_i^a$. So that the effective Schr\"{o}dinger 
equation (\ref{schreff}) contains a positive centrifugal potential 
and a Coulomb type potential with
the oscillating coefficient function $\pro_r \psi V_4[f]$.
It is known that a quantum mechanical problem for a particle in
the potential well with small enough depth does not admit bound states
(in the space of dimension $d\geq 3$). 
Therefore, there should be a critical value 
of the amplitude $A_0$ of the monopole solution, (\ref{asymsol}),
below which the eigenvalue spectrum becomes positive. 
Indeed, numerical study of solutions to the Eq.(\ref{schreff}) 
implies a critical value $a_{\rm 1cr}^{\rm bound}=1.3$. 
The monopole field $\psi(r,t)$ is
approximated by a simple interpolating function
\be
\psi^{\rm int}=1-\dfrac{(1-a_0)r^2}{1 + r^2} 
 +A_0 (1 -{\rm e}^{-d_0 r^2}) \cos (M r+b_0) \sin(M t), \quad \label{interpolfun}
\ee
where $d_0$ and $b_0$ are fitting parameters.
A typical field configuration of $f_i^a$ corresponding
to the background monopole field  $\psi(r,t)$
with the asymptotic parameters  $a_0=0.895, A_0=0.615$
and $g=1, M=1$ is shown in Fig.2.

\begin{figure}[h!]
\centering
\includegraphics[width=77mm,height=62mm,bb=0 0 473 364]
{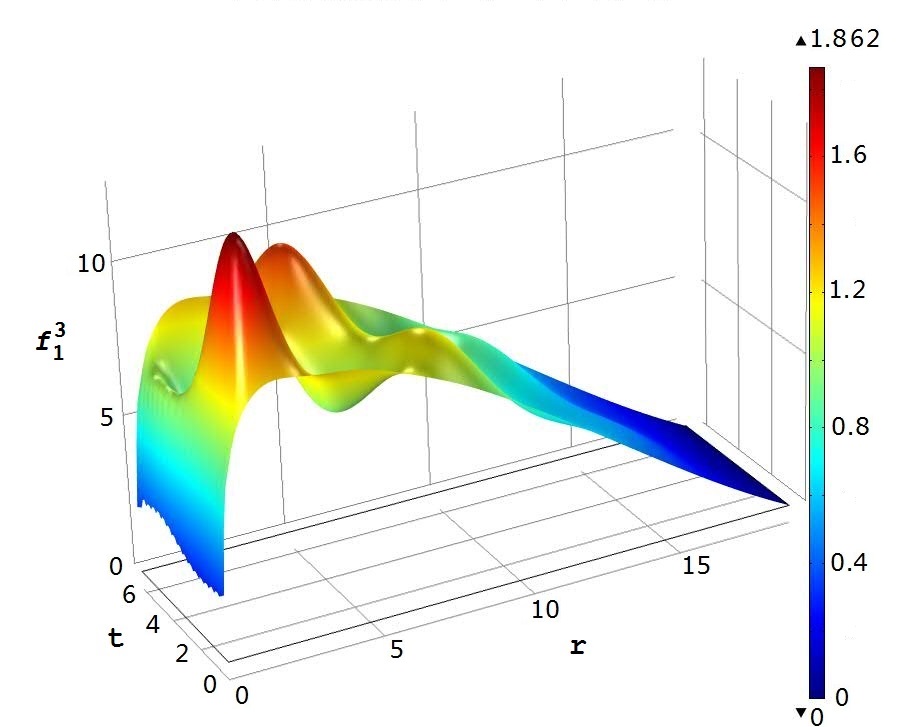}
\caption[fig2]{Wave function $f_1^3$ for the ground state with eigenvalue
$\lambda_0=0.0293$, $L=6 \pi$.}\label{Fig2}
\end{figure}

\subsection{Exact numeric solution to eigenvalue problem}
  
The qualitative estimates obtained in the previous subsection provide
only an upper bound for the critical parameter $a_{\rm 1cr}$. 
To prove rigorously that there is no negative eigenvalues 
one should solve numerically the original Schr\"{o}dinger type 
equation (\ref{schr3}) without any approximation
on functional dependence of the eigenfunctions describing the gluon fluctuations.
The equation  (\ref{schr3})  contains nine non-linear partial
differential equations which should be solved on three-dimensional 
numeric domain with a sufficient numeric accuracy.
An additional technical difficulty in numeric calculation 
on three-dimensional space-time is that one must solve
the equations with changing
the size of the numeric domain in radial direction in the limit $r \rightarrow \infty$
to verify that all eigenvalues remain positive. Fortunately, the numeric analysis
of the solutions corresponding to the lowest eigenvalue is simplified drastically
due to the factorization property of the original equation  (\ref{schr3})
and special features of the ground state solutions as we will see below.

The eigenvalue equation (\ref{schr3}) written in the component 
form admits factorization,
and it can be rewritten as two decoupled systems of partial differential equations
as follows (for brevity of notation we set $g=1$ since the coupling constant
can be absorbed by the monopole function $\psi(r,t)$)\\
(I):
\vspace*{-3mm}
\bea
&&(\hat\Delta \Psi)_2^2-\dfrac{2}{r^2}\pro_\theta \Psi_1^2
+ \dfrac{1}{r^2}\Big ((\psi^2-1)\Psi_2^2-2\psi^2 \Psi_3^1+
2 \csc^2\theta (\Psi_2^2+\Psi_3^1)+2\cot \theta \psi \Psi_3^3
 \Big )
=\lambda \Psi_2^2 , \nn \\
&&(\hat\Delta \Psi)_3^1-\dfrac{2}{r^2}\psi \pro_\theta \Psi_3^3
+
\dfrac{1}{r^2}\Big (\psi^2(-2\Psi_2^2+\Psi_3^1)-\Psi_3^1+2\csc^2\theta(\Psi_2^2+\Psi_3^1)+2\cot \theta\Psi_1^2
   \Big ) =\lambda \Psi_3^1 , \nn \\
&&(\hat\Delta \Psi)_1^2+\dfrac{2}{r^2}\pro_\theta \Psi_2^2
+\dfrac{1}{r^2}\Big ((\cot^2 \theta + \psi^2)\Psi_1^2
+2\cot \theta (\Psi_2^2+\Psi_3^1) +2  \psi \Psi_3^3 +2\Psi_1^2   \Big )
\nn \\
&&\qquad-\dfrac{2}{r} \pro_r \psi \Psi_3^3=\lambda \Psi_1^2 ,  
\label{eqI}
\\
&&(\hat\Delta \Psi)_3^3+\dfrac{2}{r^2}\psi \pro_\theta \Psi_3^1
+
\dfrac{1}{r^2}\Big (2\psi \Psi_1^2+2  \cot \theta \psi (\Psi_2^2+\Psi_3^1)
+2  \psi^2 \Psi_3^3 +\csc^2\theta \Psi_3^3\Big )
\nn \\
&& 
\qquad -\dfrac{2}{r}\pro_r \psi \Psi_1^2=\lambda \Psi_3^3 ,\nn 
\eea
(II):
\vspace*{-3mm}
\bea
&& (\hat\Delta \Psi)_1^1+\dfrac{2}{r^2}\pro_\theta \Psi_2^1-\dfrac{2}{r^2}\psi \pro_\theta \Psi_1^3
 +\dfrac{1}{r^2}\Big ((2+\cot^2 \theta +\psi^2) \Psi_1^1+2\psi \Psi_2^3
-2\cot \theta (\Psi_3^2-\Psi_2^1) \Big )\nn \\
&&\qquad -\dfrac{2}{r}\pro_r \psi \Psi_2^3
=\lambda \Psi_1^1 , \nn \\
&&(\hat\Delta \Psi)_2^3-\dfrac{2}{r^2}\pro_\theta \Psi_1^3+\dfrac{2}{r^2}\psi\pro_\theta \Psi_2^1
+\dfrac{1}{r^2}\Big (2\psi \Psi_1^1
+2\cot \theta \psi (\Psi_2^1-\Psi_3^2)+(2 \psi^2 +\csc^2\theta) \Psi_2^3  \Big )\nn \\
&&\qquad -\dfrac{2}{r}\pro_r \psi \Psi_1^1=\lambda \Psi_2^3 , \nn \\
&&(\hat\Delta \Psi)_2^1-\dfrac{2}{r^2}\pro_\theta \Psi_1^1-\dfrac{2}{r^2}\psi\pro_\theta \Psi_2^3
+\dfrac{1}{r^2}\Big (-2\psi\Psi_1^3
+ \psi^2(\Psi_2^1+2 \Psi_3^2) +2\csc^2\theta (\Psi_2^1-\Psi_3^2)-\Psi_2^1  \Big )\nn\\
&&\qquad +\dfrac{2}{r}\pro_r \psi \Psi_1^3
=\lambda \Psi_2^1 ,  
\label{eqII}
\\
&&(\hat\Delta \Psi)_1^3+\dfrac{2}{r^2}\pro_\theta \Psi_2^3+\dfrac{2}{r^2}\psi\pro_\theta \Psi_1^1
+\dfrac{1}{r^2}\Big ( 2\cot \theta\psi \Psi_1^1+2(1+\psi^2) \Psi_1^3
-2  \psi(\Psi_2^1+\Psi_3^2) \nn \\
&&\qquad +2\cot\theta\Psi_2^3 \Big ) +
\dfrac{2}{r}\pro_r \psi(\Psi_2^1+ \Psi_3^2)
=\lambda \Psi_1^3 , \nn \\
&&(\hat\Delta \Psi)_3^2
+\dfrac{1}{r^2}\Big (2\cot \theta (\psi \Psi_2^3-\Psi_1^1)
-2\psi\Psi_1^3+\psi^2(2\Psi_2^1+\Psi_3^2)-2\csc^2\theta (\Psi_2^1-\Psi_3^2)
- \Psi_3^2  \Big )\nn \\
&&\qquad +\dfrac{2}{r}\pro_r \psi \Psi_1^3=\lambda \Psi_3^2 , \nn
\eea
where
\bea
\hat \Delta \Psi_m^a \equiv -(\pro^2_{t}+\pro^2_r+\dfrac{2}{r}\pro_r+\dfrac{1}{r^2}\pro^2_\theta
+\dfrac{\cot\theta}{r^2}\pro_\theta)\Psi_m^a.
\eea

To solve numerically the systems of equations (I), (II)
 we choose a rectangular three-dimensional domain
$(0\leq t \leq 2 \pi, r_0\leq r \leq L, 0\leq \theta \leq \pi )$
and use the same interpolating function for the monopole solution $\psi(r,t)$
as before. An obtained numeric solution to the system of equations 
(I), $(\ref{eqI})$, 
implies that the lowest eigenvalue is positive,
$\lambda_{\rm I}=0.0531$, and the corresponding 
eigenfunctions have the following properties:
the functions $\Psi_1^2$ and $\Psi_3^3$ vanish identically, 
and the remaining two  functions
are related by the constraint $\Psi_3^1=-\Psi_2^2$. So that there is only one
independent non-vanishing function which can be chosen as $\Psi_2^2$.
An important feature of the solution corresponding to the lowest eigenvalue
is that the functions $\Psi_3^1,\Psi_2^2$ do not depend on the polar angle, Fig.3.
\begin{figure}[!h]
\centering
\includegraphics[width=77mm,height=62mm,bb=0 0 473 364]{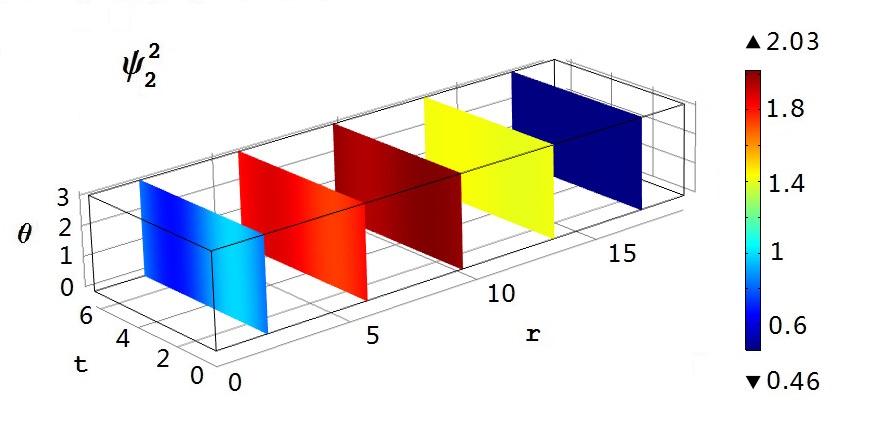}
\caption[fig3]{Eigenfunction 
$\Psi_2^2$ for the ground state with the lowest eigenvalue
$\lambda_I=0.0531$, $a_0=0.895,~A_0=0.615$, $ 0\leq r \leq 6 \pi$
,~ $0\leq t \leq 2 \pi $,
$0\leq\theta \leq \pi$.}\label{Fig3}
\end{figure}
This allows to simplify the system of equations (I)
in the case of ground state solutions with the lowest eigenvalues.
One can easily verify that the system of equations (I), (\ref{eqI}),
reduces to one partial differential equation on two-dimensional space-time
\be
\Big (-\pro^2_{t}-\pro^2_{r}-\dfrac{2}{r}\pro_r 
+\dfrac{1}{r^2}(3\psi^2-1) \Big) \Psi_2^2
=\lambda \Psi_2^2.
\ee
The last equation represents a simple Schr\"{o}dinger type equation
for a quantum mechanical problem. 
The equation does not admit negative eigenvalues if the parameter
$a_0$ of the monopole solution satisfies
the condition $a_0\geq 1/\sqrt 3\simeq 0.577\cdots$
which provides a totally repulsive quantum mechanical potential in this equation.

A structure of the system of equations (II), (\ref{eqII}), admits 
a similar factorization property
on the space of ground state solutions.
We have solved numerically the equations (II)
with the same background monopole function $\psi(r,t)$ for various values
of the parameters $a_0, A_0, M$. In a special case, 
$a_0=0.895,~A_0=0.615$, $g=1,M=1$, $ 0\leq r \leq 6 \pi$
the obtained numeric solution for the ground state has a lowest
 eigenvalue $\lambda_{\rm II}=0.0142$ which is less than $\lambda_{\rm I}$.
 All components of the solution do not have dependence on the polar angle and
satisfy the following relationships: $\Psi_2^1=\Psi_3^2$ and
$\Psi_1^1=\Psi_2^3=0$. There are two independent non-vanishing functions
which can be chosen as $\Psi_1^3$ and $\Psi_3^2$.
One can check that on the space of solutions corresponding to 
the lowest eigenvalue the system of equations (II) is reduced
to two coupled partial differential equations for two
functions $\Psi_1^3(r,t)$ and $\Psi_3^2(r,t)$, 
\be
\renewcommand{\arraystretch}{1.8}
\begin{array}{rcl}
(-\pro^2_{t}-\pro^2_{r}-\dfrac{2}{r}\pro_r ) \Psi_1^3
+\dfrac{2}{r^2}\Big( (1+\psi^2)\Psi_1^3-2 \psi \Psi_3^2
\Big ) 
+\dfrac{4}{r} \pro_r\psi \Psi_3^2&=&\lambda \Psi_1^3, \\
(-\pro^2_{t}-\pro^2_{r}-\dfrac{2}{r}\pro_r ) \Psi_3^2
+\dfrac{1}{r^2} \Big ((3 \psi^2-1) \Psi_3^2-2 \psi \Psi_1^3
\Big ) 
+\dfrac{2}{r}\pro_r \psi \Psi_1^3&=&\lambda \Psi_3^2. 
\end{array}
\label{sys2}
\renewcommand{\arraystretch}{1.0}
\ee
Exact numeric solution profiles for the functions $\Psi_1^3, \Psi_3^2$ are shown in
 Fig.4. 
\begin{figure*}[!h]
  \subfigure[~ $\Psi_1^3$]
{\includegraphics[scale=0.38,bb=0 0 498 359]{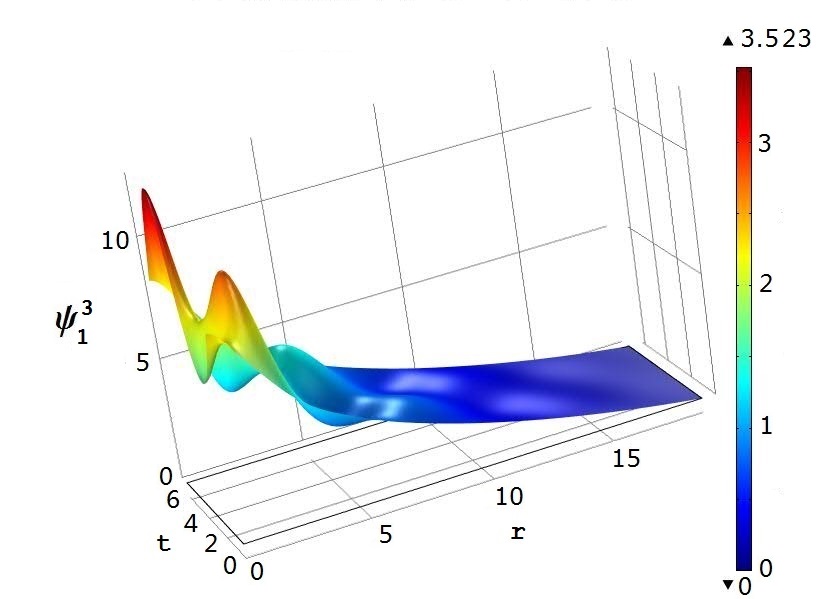}}
\qquad\qquad\qquad
  \subfigure[~ $\Psi_3^2$]
{\includegraphics[scale=0.38,bb=0 0 498 359]{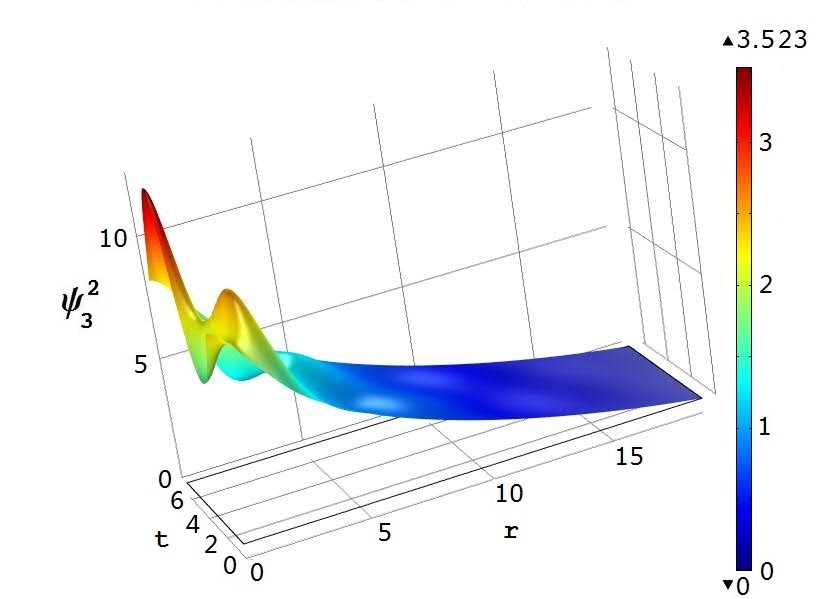}}
\caption[fig4]{Solution profiles for the functions
$\Psi_1^3(r,t)$, (a), $\Psi_3^2(r,t)$, 
(b), corresponding to the lowest eigenvalue $\lambda =0.014218$,
$a_0=0.895,~A_0=0.615$,
$ 0\leq r \leq 6 \pi$,~ $0\leq t \leq 2 \pi $.}\label{Fig4}
\end{figure*}
We have obtained that the lowest eigenvalue is positive
when asymptotic monopole amplitude is less than
a critical value $a_{\rm 1cr} \simeq 0.625$.
We conclude that since the system of equations (II) has ground state solution 
with the lower eigenvalue than ground state solution to the system (I), 
it is enough to study the properties of the couple of 
equations (\ref{sys2}) defined on two-dimensional space-time.

Note that numeric solving of the original eigenvalue equations (\ref{schr3})
on a three-dimensional domain does not provide high numeric accuracy,
especially in the limit of large values of the size $L$ of the numeric 
domain along the radial direction.
Remind that one should prove the positiveness of the ground state 
eigenvalues in the limit of infinite space
when the eigenvalues become very close to zero.
Solving the reduced equations (\ref{sys2}) defined on two-dimensional space-time
can be performed easily using standard numeric packages
with sufficient accuracy. 
The obtained numeric accuracy for the eigenvalues $\lambda(L)$ 
in solving the two-dimensional equations  (\ref{sys2})
is $1.0 \times 10^{-5}$ which is enough to construct the
eigenvalue dependence on the size $L$ of the box in
the range  $6\pi\leq L \leq 64 \pi$.

\section{Vacuum stability of the monopole field in $SU(3)$ QCD}

The analysis of quantum stability 
 of the stationary monopole solution performed in the previous section
 can be generalized straightforward to the case of 
a pure $SU(3)$ QCD. 
In general the  system of two partial differential equations (\ref{eqsu3}) 
admits non-stationary and quasi-stationary solutions.
We are interested in stationary monopole solutions
which can be obtained in a full analogy with the case of $SU(2)$
QCD by using a constraint $\psi_1=\psi_2 \equiv \tilde \psi$.
The Schr\"{o}dinger type equation (\ref{schr3}) for possible 
unstable modes corresponding to quantum gluon fluctuations contains twenty four
partial differential equations. Since the initial ansatz (\ref{spherwav})
describes monopole solution corresponding to $I$- and $U$-type subgroups 
$SU(2)$ one expects that factorization property and
reduction of the equations on the space of ground state solutions
will take place in a similar manner as in the case of $SU(2)$ QCD.
Indeed, the numeric analysis of ground state solutions 
of the full $SU(3)$  Schr\"{o}dinger equation (\ref{schr3}) 
with charge two monopole background
field $\tilde \psi$ implies factorization of the equations in each 
sector of $I,U$ subgroups $SU(2)$.
The obtained numeric solution with the lowest eigenvalue
has a basis which contains six non-vanishing functions $\tPsi_1^3,\tPsi_3^2,
\tPsi_2^1= \tPsi_2^6  =\tPsi_3^7=\tPsi_3^2,\tPsi_1^8=\sqrt 3 \tPsi_1^3$,
all other functions are vanished identically.
An important feature of the numeric solution is that
all solution profile functions do not depend on the polar angle $\theta$. 
One can  choose the functions
 $\tPsi_1^3,\tPsi_3^2$ as independent ones and verify that twenty four 
 equations in the Schr\"{o}dinger type equation (\ref{schr3}) reduce to a
 simple system of two independent partial differential equations  
 on two-dimensional space-time $(r,t)$ in a consistent manner
\be
\renewcommand{\arraystretch}{1.8}
\begin{array}{rcl}
\pro^2_r \tPsi_{1}^3+\dfrac{2}{r} \pro_r \tPsi_{1}^3+\pro_t^2\tPsi_{1}^3
-\dfrac{1}{r^2}\Big ((2 \tpsi^2+1)\tPsi_1^3-2 (\tpsi-r\tpsi_r)
\tPsi_3^2\Big )
&=&0,  \\
\pro^2_r\tPsi_{3}^2+\dfrac{2}{r}\pro_r \tPsi_{3}^2+\pro^2_t\tPsi_{3}^2 
-\dfrac{1}{2 r^2}\Big ((3\tpsi^2-2)\tPsi_3^2-4(\tpsi-r\tpsi_r) \tPsi_1^3\Big
 )&=&0.
\end{array}
\label{su3reduction}
\renewcommand{\arraystretch}{1.0}
\ee
One can show that the structure of the reduced equations 
is equivalent to the structure of the
equations (II), (\ref{sys2}), obtained by reduction in the case of $SU(2)$ QCD.
Simple rescaling of the function $\tpsi= \sqrt 2 \psi$ 
representing the monopole background field
and function renormalization $\tPsi_3^2=\sqrt 2 \Psi_3^2$
($\tPsi_1^3=\Psi_1^3$) in the Eqs.(\ref{su3reduction}) lead exactly 
to the system of equations (II),  (\ref{sys2}).

We have verified that with increasing the size of 
the space-time numeric domain in the radial direction,
  $L\rightarrow \infty$, the corresponding eigenvalue $\lambda(L)$ 
vanishes asymptotically from positive values.
The eigenvalue dependencies $\lambda(L)$ obtained 
by solving the approximate,  (\ref{schreff}), 
and exact  Schr\"{o}dinger equations, (\ref{schr3}), 
in the cases of $SU(2)$ and $SU(3$) QCD are presented in Fig.5.
\begin{figure}[h!]
\centering
\includegraphics[width=72mm,height=52mm,bb=0 0 324 201]
{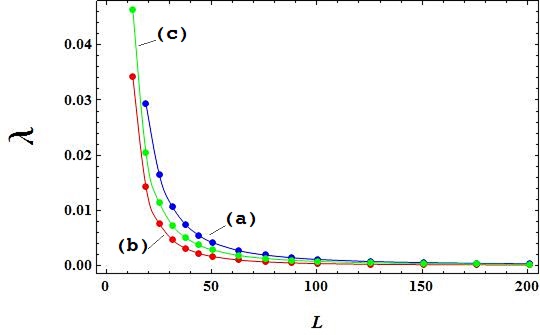}
\caption[fig5]{Lowest eigenvalue dependence, $\lambda(L)$, 
on the radial size $L$ of the numeric domain: 
(a) approximate results in $SU(2)$ QCD, 
(b) exact numeric results in $SU(2)$ QCD, 
(c) exact numeric results in $SU(3)$ QCD.}\label{Fig5}
\end{figure}

Note that in $SU(3)$ QCD one has a stable monopole field configuration
for both stationary monopole solutions, 
for the embedded $SU(2)$ monopole with a unit magnetic charge,
and for $SU(3)$ monopole with the magnetic charge two. 
The $SU(3)$ symmetric monopole solution corresponding to two subgroups $SU(2)$ 
is preferable since for constant valued functions $\psi_1, \psi_2$ the
corresponding classical potential has an absolute minimum at 
$\psi_1\!\!=\!\!\psi_2\!\!=\!\!\sqrt 2$.
This is similar to the behavior of one-loop effective potential 
for homogeneous color magnetic fields where the potential has an absolute minimum for
non-vanishing values 
of both magnetic fields $H_3$ and $H_8$ 
corresponding to Cartan subalgebra of $\mathfrak{su}(3)$
\cite{fly}. 

\section{Conclusion}

In our consideration of the stationary monopole solutions
we set the mass scale parameter $M$ to unit for simplicity.
We would like to stress the importance of the presence of such a parameter
and its relation to microscopic structure of the vacuum.
One of the main characteristics of
QCD vacuum is the vacuum gluon condensate,  $\langle \vec F_{\mu\nu}^2 \rangle$.
In first approximation the vacuum gluon condensate represents a constant parameter
which describes a mass gap in the confinement phase.
Within the framework of the confinement mechanism 
based on the monopole condensation and dual Meissner effect it is assumed that
the vacuum gluon condensate is generated due to condensation 
of monopoles, i.e., the dominant contribution to 
the gauge invariant quantity 
 $\langle\vec F_{\mu\nu}^2\rangle$ is made of color magnetic 
field.  
For the present moment 
an exact vacuum field configuration corresponding to the vacuum 
monopole condensate is unknown. We expect that QCD vacuum
is formed through the condensation of monopoles and/or monopole 
pairs since only these
objects possess  quantum stability locally in small vicinity of each
space point before any averaging procedure over the whole space-time region. 
Simple consideration shows that the mass scale parameter
$M$ determines the microscopic scale of the confinement
phase. 

Let us consider the structure of the one-loop effective potential of
a pure $SU(N)$ QCD with a color magnetic background field
which is treated as a constant field 
\cite{savv,yildiz80,claudson80,adler81,dittrich83,flory83,blau91,reuter97}
\be
V_{\rm eff}
=\dfrac{1}{4} H^2
+\dfrac{11 N g^2(\mu)}{48 \pi^2} H^2\Big(\ln \dfrac{g(\mu) }{\mu^2}
-c\, \Big),
\ee
where $g(\mu)$ is a renormalized coupling constant
defined at the subtraction point $\mu^2\simeq \Lambda_{\rm QCD}$.
The gauge invariant quantity $H^2 \equiv \langle\vec{F}^2_{\mu\nu}\rangle$ 
represents a vacuum
averaged value of the gluon field operator which describes the monopole condensate.
Note that in the standard QCD the notion of the vacuum gluon condensate 
differs from the notion of the electron pair condensate in the superconductor which
is described by a wave function of the Cooper electron pair 
(we do not consider phenomenological
approach to QCD based on Ginsburg-Landau type models).
The effective  potential has a non-trivial minimum
 at non-zero value of the vacuum magnetic field
$\langle H\rangle \simeq 0.14 \mu^2$ which fixes the scale $M$ 
of the microscopic structure
of the vacuum monopole condensate.
In the confinement phase one
assumes that the periodic structure of the classical 
monopole field configuration
is characterized by the length parameter $\lambda_M=2 \pi/M$
which should be less than the inverse deconfinement temperature parameter
$kT_{\rm dec}$, i.e.,
$2 \pi/M\leq \beta_{\rm dec}=1/kT_{\rm dec}$. 
Under this condition one can perform space-time averaging of the 
classical monopole configuration and estimate a lower bound
$M^2_{\rm bound}\simeq 1.2 \mu^2$ which is of the same order
as the parameter $\Lambda_{\rm QCD}$.
In the confinement phase, $T\simeq 0$,  the vacuum averaging value of the gluon field
operator $\langle0|A_\mu^a|0\rangle$ vanishes since the effective 
size $\beta = 1/kT$ of the time interval in the Euclidean functional integral
of the effective action becomes much larger than $\lambda_M$.
With increasing temperature the value of the parameter $\beta$
becomes comparable and less than the periodic scale $\lambda_M$
of the vacuum monopole field configuration. This implies that
the vacuum averaging value of the gluon field operator, 
$\langle0|A_\mu^a|0\rangle$, becomes
non-vanishing, which leads to the deconfinement phase with the spontaneous
symmetry breaking and gluon can be observed as a color object. 

In conclusion, 
we have demonstrated that the stationary spherically symmetric monopole
solution represents a stable quasi-classical vacuum field background 
under small gluon fluctuations.
Certainly, our consideration is restricted by one-loop consideration 
within the formalism of the effective action. 
Note that the classical monopole solution is non-perturbative
and it is valid for arbitrary values of the coupling constant. 
Besides, the one-loop effective action includes non-perturbative 
contribution due to summation over all one-loop Feynman diagrams 
corresponding to interaction of the background monopole field with virtual gluons.  
This gives a hope that  a modified monopole solution to quantum
equations of motion obtained beyond one-loop approximation 
will admit the quantum stability as well. 

Recently it has been shown that the spherically symmetric 
monopole solution is rather classically unstable under 
axially-symmetric deformations of the initial spherically-symmetric 
ansatz \cite{pakp1}.
This might cause some doubts whether spherically symmetric monopoles can serve
as a structure element of a true vacuum.
One should note that classical stability of the multi-monopole system
represents a non-trivial problem due to presence of 
interaction between the monopoles.
Another candidate for a stable structure element in formation of 
a stable vacuum has been proposed in \cite{P4} 
where it has been shown that a monopole-antimonopole pair solution
is stable under quantum gluon fluctuations. 
We expect that monopole and/or monopole-antimonopole pair 
condensation can be realized in QCD in analogy 
with the Cooper electron pair condensation in ordinary superconductor, 
as it was conjectured in the seminal papers long 
time ago \cite{nambu74, mandelstam76, polyakov77, thooft81}.
This issue will be considered in a separate paper.

\section*{Acknowledgment}

One of authors (DGP) thanks Prof. C. M. Bai for warm hospitality 
during his staying at the Chern Institute of Mathematics. 
The work is supported by National Research Foundation
of Korea, grants NRF-2014R1A2A1A01002306 and NRF-2017R1D1A1B03028310; 
Korean Federation of Science and Technology, Brain Pool Program, 
and grant OT-$\Phi$2-10.


\begin{thebibliography}{99}
\bibitem{Phil2008} O. Philipsen, 
Prog. Theor. Phys. Suppl. { \bf 174} (2008) 206.
\bibitem{diGiacomo2014} A. Di Giacomo, 
EPJ Web of Confs. {\bf 70} (2014) 00019.
\bibitem{nambu74} Y. Nambu, 
Phys. Rev. {\bf D10} (1974) 4262.
\bibitem{mandelstam76} S. Mandelstam, 
 Phys. Rep. {\bf 23} (1976) 245.
\bibitem{polyakov77} A.M.  Polyakov, 
 Nucl. Phys. {\bf B120} (1977) 429.
\bibitem{thooft81} G. 't Hooft, 
Nucl. Phys. {\bf B190} (1981) 455.
\bibitem{Nair1985} V. P. Nair and C. Rosenzweig, 
Phys. Rev. {\bf D31} (1985) 401. 
\bibitem{savv} G.K. Savvidy,   
Phys. Lett. {\bf B71} (1977) 133.
\bibitem{N-O} N.K. Nielsen and P. Olesen, 
Nucl. Phys. {\bf B144} (1978) 376.
\bibitem{niel-nino}H.B. Nielsen and M. Ninomiya, 
Nucl. Phys. {\bf B156} (1979) 1.
\bibitem{niel-oles} H.B. Nielsen and P. Olesen, 
Nucl. Phys. {\bf B160} (1979) 380.
\bibitem{amb-oles1} J. Ambj{\o}rn and P. Olesen,
 Nucl. Phys. {\bf B170} (1980) 60.
\bibitem{shuryak1997} E. Shuryak and T. Schafer, 
Annual Review of Nuclear and Particle Science, {\bf 47} (1997) 359.
\bibitem{1995simonov} Yu.A.Simonov, 
Lecture at the International School of Physics ``Enrico Fermi'', 
Varenna, 27 June--7 July 1995, {\tt [arXiv:hep-ph/09509403]}. 
\bibitem{kumar2010} B. S. Rajput and S. Kumar, 
Advances in High Energy Physics, Vol. 2010 (2010) 713659. 
\bibitem{engelhardt2000} M. Engelhardt, K. Langfeld, H. Reinhardt, 
and O. Tennert, 
Phys. Rev. {\bf D61} (2000) 054504.
\bibitem{choprl80} Y.M. Cho,   
Phys. Rev. Lett. {\bf 44} (1980) 1115.
\bibitem{chopakprd2002} Y.M. Cho and D.G. Pak, 
Phys. Rev. {\bf D65} (2002) 074027.
\bibitem{diGiacomo2015} A. Di Giacomo, 
Talk at the Conf. QCD-2015, Montpellier (France), June 29- July 4
(2015),  \\ 
{\tt [arXiv:1509.02288[hep-lat]]}. 
\bibitem{pak05} Y.M. Cho and D.G. Pak, 
 Phys. Lett. {\bf B632} (2006) 745.
\bibitem{schan82} V. Schanbacher, 
Phys. Rev. {\bf D26} (1982) 489.
\bibitem{bordag} M. Bordag, 
Phys. Rev. {\bf D67} (2003)  065001.
\bibitem{pakp1} B.-H. Lee, Y. Kim, D.G. Pak, T. Tsukioka, and P.M. Zhang,
 Int. J. Mod. Phys. {\bf A32} (2017) 1750062. 
\bibitem{p14} M. Luscher,
Phys. Lett. {\bf B70} (1977) 321.
\bibitem{p15} B. Schechter,   
Phys. Rev. {\bf D16} (1977) 3015.
\bibitem{p16} H. Arodz, 
Phys. Rev. {\bf D27} (1983) 1903.
\bibitem{p17} E. Farhi, V.V. Khoze, and R Singleton,   
Phys. Rev. {\bf D47} (1993) 5551.
\bibitem{p18} A. Abouelsaood and M.H. Emam, 
Phys. Lett. {\bf B412} (1997) 328.
\bibitem{choprd80} Y. M. Cho, 
Phys. Rev. {\bf D21} (1980) 1080.
\bibitem{derr} G. H. Derrick, 
J. Math. Phys. {\bf 5} (1964) 1252.
\bibitem{jackiw77} R. Jackiw, 
The Yang-Mills Vacuum as a Bloch Wave, preprint MIT-CTP-625, (MIT, LNS), 
(1977) 12 pp.
\bibitem{jackiwRMP} 
R. Jackiw, 
Rev. Mod. Phys. {\bf 49} (1977) 681. 
\bibitem{yildiz80} A. Yildiz and P. Cox, 
Phys. Rev. {\bf D21} (1980) 1095.
\bibitem{claudson80}  M. Claudson, A. Yilditz, and P. Cox,
    Phys. Rev. {\bf D22} (1980) 2022.
\bibitem{adler81}  S. Adler,   
Phys. Rev. {\bf D23} (1981) 2905.
\bibitem{dittrich83} W. Dittrich and M. Reuter, 
Phys. Lett. {\bf B128} (1983) 321.
\bibitem{flory83}  C. Flory, 
Phys. Rev. {\bf D28} (1983) 1425.
\bibitem{blau91}  S.K. Blau, M. Visser, and A. Wipf, 
 Int. J. Mod. Phys. {\bf A6} (1991) 5409.
\bibitem{reuter97}  M. Reuter, M.G. Schmidt and C. Schubert,
Ann. Phys. {\bf 259} (1997) 313.
\bibitem{abbott} L.F. Abbott, 
Acta Phys. Polon. {\bf B13} (1982) 33.
\bibitem{siegel}  M.T. Grisaru and W. Siegel, 
Nucl. Phys. {\bf B187} (1981) 149.
\bibitem{fly} H. Flyvbjerg,   
Nucl. Phys. {\bf B176} (1980) 379.
\bibitem{P4}  D.G. Pak, B.-H. Lee, Y. Kim, T. Tsukioka, and P.M. Zhang
{\tt[arXiv:1703.09635[hep-th]]}.
\end{thebibliography}
\end{document}